\documentclass[10pt,journal,twocolumn]{IEEEtran}
\usepackage{graphicx}
\usepackage{dcolumn}
\usepackage{amsmath,amsfonts}
\usepackage{bm}
\usepackage{multirow}
\usepackage{slashbox}
\usepackage{tabularx}
\usepackage{tikz}
\usepackage{mathtools,amssymb}
\begin{document}
\title{Coupling a generative model with a discriminative learning framework for speaker verification}
\author{Xugang Lu$^{1}$, Peng Shen$^{1}$, Yu Tsao$^{2}$, Hisashi Kawai$^1$
\thanks{1. National Institute of Information and Communications
Technology, Japan. Email: xugang.lu@nict.go.jp
}
\thanks{2. Research Center for Information Technology Innovation, Academic
Sinica, Taiwan} }

\IEEEcompsoctitleabstractindextext{%
\begin{abstract}
\small{}
The task of speaker verification (SV) is to decide whether an utterance is spoken by a target or an imposter speaker. In most studies of SV, a log-likelihood ratio (LLR) score is estimated based on a generative probability model on speaker features, and compared with a threshold for making a decision. However, the generative model usually focuses on individual feature distributions, does not have the discriminative feature selection ability, and is easy to be distracted by nuisance features. The SV, as a hypothesis test, could be formulated as a binary discrimination task where neural network based discriminative learning could be applied. In discriminative learning, the nuisance features could be removed with the help of label supervision. However, discriminative learning pays more attention to classification boundaries, and is prone to overfitting to a training set which may result in bad generalization on a test set. In this paper, we propose a hybrid learning framework, i.e., coupling a joint Bayesian (JB) generative model structure and parameters with a neural discriminative learning framework for SV. In the hybrid framework, a two-branch Siamese neural network is built with dense layers that are coupled with factorized affine transforms as used in the JB model. The LLR score estimation in the JB model is formulated according to the distance metric in the discriminative learning framework. By initializing the two-branch neural network with the generatively learned model parameters of the JB model, we further train the model parameters with the pairwise samples as a binary discrimination task. Moreover, a direct evaluation metric (DEM) in SV based on minimum empirical Bayes risk (EBR) is designed and integrated as an objective function in the discriminative learning. We carried out SV experiments on Speakers in the wild (SITW) and Voxceleb. Experimental results showed that our proposed model improved the performance with a large margin compared with state of the art models for SV.
\end{abstract}
}

\maketitle

\IEEEdisplaynotcompsoctitleabstractindextext

\IEEEpeerreviewmaketitle

\section{Introduction}
\label{sec-I}
Speaker verification (SV) is a technique to verify whether an acoustic speech is spoken by a target or an imposter speaker. SV is widely used in many speech application systems where speaker information is required from authentication or security perspectives \cite{Hansen2015,Beigi2007,Dehak2011}. The basic problem definition of SV is to decide whether two utterances (usually denoted as test and enrollment utterances) are generated from the same or different speakers, i.e., a hypothesis test defined as:
\begin{equation}
\begin{array}{l}
  H_S :{\bf x}_i ,{\bf x}_j \mathop {}\limits^{} \text{are spoken by the same speaker} \\
 H_D :{\bf x}_i ,{\bf x}_j \mathop {}\limits^{} \text{are spoken by different speakers} \\
 \end{array}
 \label{eq_HSD}
 \end{equation}
where $H_S$ and $H_D$ are the two hypotheses as the same and different speaker spaces, respectively. $({\bf x}_i ,{\bf x}_j$) is a tuple with two compared utterances indexed by $i$ and $j$.  For making a decision, it is necessary to estimate the similarity of the two utterances, either calculated as a log likelihood ratio (LLR) or a distance metric measure, and compare it with a threshold. The conventional pipeline in constructing a SV system for doing the hypothesis test defined in Eq. (\ref{eq_HSD}) is composed of front-end speaker feature extraction and backend speaker classifier modeling. Front-end feature extraction tries to extract robust and discriminative features to represent speakers, and backend classifier tries to model speakers with the extracted features based on which the similarity or LLR scores could be estimated.
\subsection{Front-end speaker feature extraction}
Historically, in most state of the art frameworks, the front-end speaker feature was based on i-vector representation \cite{Dehak2011}. In i-vector extraction, speech utterances with variable durations can be converted to fixed dimension vectors with the help of Gaussian mixture models (GMM) on probability distributions of acoustic features. With the resurgence of deep learning techniques, several alternative speaker features have been proposed, e.g., d-vector \cite{Variani2014} and X-vector \cite{Snyder2018}.  These features are extracted from a well trained deep neural network with bottleneck layers or statistical pooling. In recent years, X-vector as one of the speaker embedding representations is widely used in most state of the art frameworks \cite{Snyder2018}. The advantage of X-vector representation is that the model for X-vector extraction could be efficiently trained with a large quantity of speech samples from various speakers. Moreover, in order to explore robust speaker information, data augmentation with various noise types and signal to noise ratios (SNRs) could be easily applied in model training \cite{Snyder2018}. Since the original front-end feature (e.g., either i-vector or X-vector) encodes various acoustic factors, e.g., speaker factor, channel transmission factor, recording device factor, etc., before classifier modeling, a linear discriminative analysis (LDA), or a local fisher discriminative analysis \cite{SugiyamaFLDA,ShenICASSP2016}, is usually applied for dimension reduction to eliminate non-speaker specific information.
\subsection{Backend classifier modeling}
\label{sub_backend}
After speaker features are obtained, how to build a speaker classifier model in backend modeling for SV is important. There are two types of modeling approaches, one is generative modelling, the other is discriminative modeling. In generative modeling, features are regarded as observations from a generation process with certain probability distribution assumptions on the generation variables. Based on the generation model, the hypothesis test defined in Eq. (\ref{eq_HSD}) is regarded as a statistical inference from the variable probability distributions. For example, probabilistic linear discriminant analysis (PLDA) modeling was originally proposed for face recognition in \cite{Prince2007}, and was later improved with many variants for biometric authentication \cite{PLDAUni}. It has been widely used in SV for building classifier or backend models \cite{Kenny2010,Kenny2013}. PLDA can be applied to model the within-speaker and between-speaker variabilities with linear subspace modeling on speaker and noise spaces in generation. However, it is difficult to determine the dimensions of subspaces, which has a large effect on the final performance. As an alternative, joint Bayesian (JB) modelling \cite{ChenJB2012,ChenJB2016}, which is without subspace model assumptions on speaker and noise spaces, is regarded as a much more efficient model than PLDA. Besides using different modelling assumption from that used in PLDA, JB has a quick convergence speed and accuracy in model parameter estimation with expectation-maximization (EM) iterations \cite{ChenJB2012,ChenJB2016, WangOu2017}. The other approach in backend modeling is discriminative modeling. In the early stage, cosine distance metric as a measure of similarity between two compared speaker embedding features was widely used \cite{Dehak2011}. With proper speaker feature extractions, the performance based on cosine distance may outperform the PLDA based backend modeling \cite{Vox2020Web}. However, the scores estimated based on cosine distance metric need a lot of careful post processing, for example, score normalization or imposter cohort selection. For unknown or unconstraint environments or conditions, the generative probabilistic models are much more suitable for capturing the latent variations of the acoustic environments. Since the hypothesis test defined in Eq. (\ref{eq_HSD}) also can be formulated as a binary classification task, a discriminative modeling approach can be applied with supervised learning algorithms. For example, support vector machines (SVM) were proposed to maximize the between class distance \cite{WANSVM2000,SVMSV3}, and a neural network based discriminative model was applied to directly maximize classification accuracy with labeled training data sets \cite{Villalba2017}.  The first explicit discriminative training based on pairwise i-vector features was proposed as a binary classification task for SV in \cite{Burget2011}, and later the idea was further developed to connect the PLDA based scoring to kernel classifier in pairwise i-vector space \cite{Cumanni2013}. In recent years, as a discriminative modeling approach, the supervised end-to-end speaker models which integrate the front-end feature extraction and backend speaker classifier modeling in a unified optimization framework also have been proposed \cite{Heigold2016,WanLi2018}. However, in SV tasks, usually many speakers are not registered in the training data, and test utterances may be recorded from different sessions and environments, so it is difficult for the supervised discriminative modeling to work well if the training and testing conditions are not matched. To deal with the unmatched conditions, several backend algorithms have been proposed \cite{backend1,backend2}. No matter how successful the neural network based discriminative modeling in speech and image, current state of the art pipeline for SV is still the speaker embedding feature extraction (e.g. X-vector) with a generative speaker classifier modeling.
\subsection{Our focus: hybrid generative and discriminative approach for backend modeling}
\begin{figure}[tb]
\centering
\includegraphics[scale=1]{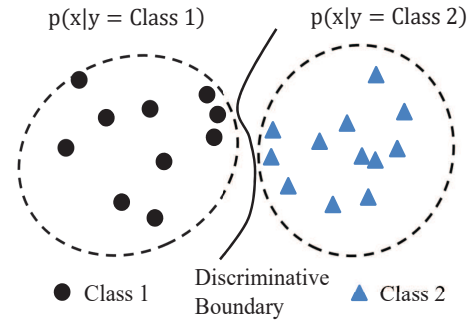}
\caption{Generative model learning focuses on class conditional feature distributions (dashed-circles of feature distribution shapes), and discriminative model learning emphasizes the class discriminative boundary (solid curve).}
\label{fig_GD}
\end{figure}

In this study, we focus on backend modeling in SV tasks. We first summarize the different focus of the generative and discriminative modeling approaches as reviewed in section \ref{sub_backend}. For a better understanding, a two-class classification task is illustrated in Fig. \ref{fig_GD} (circles and triangles for classes 1 and 2 respectively). As shown in this figure, the generative model tries to focus on class conditional feature distributions while the discriminative model tries to pay attention to the classification boundary (solid curve in Fig. \ref{fig_GD}). As a generative modelling approach, either with PLDA or JB models, prior probability distributions of variables are assumed. If the assumptions are not satisfied, the performance could not be guaranteed. Moreover, it is difficult for the generative model approach to learn data structures in a high dimensional space with complex distributions. And the model does not have the discriminative feature selection ability which may be easily distracted by nuisance features in learning. On the other hand, the discriminative model approach could learn the complex classification boundary with a strong ability to remove nuisance features, but is prone to overfitting to the training data with over estimated label confidence in training.
In this study, we try to explicitly integrate both the advantages of generative and discriminative modeling approaches in a unified learning framework for SV. The idea of discriminative training with a generative model scoring is not new \cite{Burget2011,Cumanni2013, Rohdin2020, neuralPLDA}, the novelty of our work lies in the way of how to exactly coupling the generative model parameters in a discriminative learning framework. Besides, after coupling the generative model parameters in a discriminative learning framework, direct evaluation metrics could be designed as learning objective functions. Our contributions are summarized as follows:

(1) We propose a unified neural network backend framework for SV which couples the JB based generative model parameters in a discriminative learning framework. Although hybrid generative and discriminative modelling has been studied in machine learning for fully utilizing labeled and unlabeled samples, and showed improved performance in classification tasks \cite{Lasserre2006}, it is difficult to integrate the generative and discriminative models in SV tasks. The main reason is that in most studies the generative and discriminative models adopted different modeling structures. In this study, we take the matrix structure of the generative JB model into consideration during the design of a neural network based discriminative modeling framework.

(2) We design a direct evaluation metric based learning objective function which keeps consistency of using the same evaluation metric in both training and testing. In the JB based generative model learning, usually an objective function with negative log-likelihood is minimized, while in a neural network based discriminative model learning, an objective function indicating the classification error rate is minimized. However, the objective for the hypothesis test in SV is different from any of them. In a SV task, the evaluation metric is based on weighting two types of errors (miss and false alarm) \cite{BOSARIS2011,Lehmann2005}. In this study, we formulate this type of objective function in the discriminative learning framework.

(3) We analyze the effects of all components in model structure and parameterizations with detailed SV experiments, and reveal their connections to conventional distance metric learning.

The remainder of the paper is organized as follows. Section \ref{sec-II} introduces the basic theoretical considerations and the proposed hybrid model framework. Section \ref{sec-III} describes the implementation details and experiments; in particular, we make deep investigations of the effect of model parameters, and their connections to other related model frameworks. Section \ref{sec-IV} further checks the effectiveness of the proposed framework on another advanced speaker feature extraction baseline. Conclusions and future work are given in Section \ref{sec-conclud}.

\section{Proposed hybrid model framework}
\label{sec-II}
The generative and discriminative models can be connected with the Bayes theory. Before introducing their connections, we give a brief review of generative and discriminative models.
\subsection{Generative and discriminative models in classification tasks}
A generative model tries to capture the data generation process with a fully joint modelling of the relation between feature input and label variables as $p\left( {{\bf x},{y}} \right)$, while a discriminative model only tries to model the direct relation between input feature and output label as $p\left( {y|{\bf x}} \right)$, where $\bf x$ and $y$ are feature and label variables, respectively. Although the generative model is not directly used for classification, a classification model can be deduced from the generative model as model inference based on the Bayes theory as:
\begin{equation}
p\left( {y|{\bf x}} \right) = \frac{{p\left( {{\bf x},y} \right)}}{{p\left( {\bf x} \right)}} = \frac{{p\left( {{\bf x}|y} \right)p\left( y \right)}}{{p\left( {\bf x} \right)}},
\label{eq_Bayes}
\end{equation}
where ${p\left( {{\bf x}|y} \right)}$ is the likelihood score of generating feature $\bf x$ by given a label $y$. In practical model parameter learning,  generative model parameters usually are estimated based on expectation-maximization (EM) like algorithms while discriminative model parameters (neural network) usually are estimated based on gradient descent algorithms. In the following subsections, we show how to integrate them in a hybrid model with careful formulations.
\subsubsection{Generative model based classification}
Given a training data set $\left\{ {\left( {{\bf x}_i ,y_i } \right)} \right\}_{i = 1,2,...,N} ,y_i  \in \left\{ {1,2,...,K} \right\}$ with ${\bf x}_i$ and $y_i $ as data feature and label, and $K$ the number of classes, for a classification based on a generative model, based on Eq. (\ref{eq_Bayes}), the classification model is:
\begin{equation}
p\left( {y = k|{\bf x}} \right) = \frac{{p\left( {{\bf x}|y = k} \right)p\left( {y = k} \right)}}{{\sum\limits_{j = 1}^K {p\left( {{\bf x}|y = j} \right)p\left( {y = j} \right)} }}.
\label{eq_Gen1}
\end{equation}
And Eq. (\ref{eq_Gen1}) is further cast to:
\begin{equation}
p\left( {y = k|{\bf x}} \right) = \frac{1}{{1 + \sum\limits_{j = 1,j \ne k}^K {\exp \left( {-r_{k,j} \left( {{\bf x},\Theta _G } \right)} \right)} }},
\label{eq_Gen2}
\end{equation}
where
\begin{equation}
r_{k,j} \left( {{\bf x},\Theta _G } \right) = \log \frac{{p\left( {{\bf x}|y = k} \right)p\left( {y =k} \right)}}{{p\left( {{\bf x}|y = j} \right)p\left( {y = j} \right)}},
\label{eq_GenG}
\end{equation}
is a LLR score based on class generative probability model with $\Theta _G$ as model parameter set.
\subsubsection{Discriminative model based classification}
Rather than using a generative model, a neural network can be applied to directly approximate the posterior probability function $p\left( {y|{\bf x}} \right)$. A discriminative learning tries to approximate the mapping between input feature and label with a softmax function defined as:
\begin{equation}
p\left( {y = k|{\bf x}} \right) = \frac{{\exp \left( {o_k } \right)}}{{\sum\limits_{j = 1}^K {\exp \left( {o_j } \right)} }},
\label{eq_softmax}
\end{equation}
where a network mapping function $o_j  = \phi _j \left( {{\bf x},\Theta_D } \right)$ is defined as the output corresponding to the $j$-th class, and $\Theta_D$ is the neural network parameters. Eq. (\ref{eq_softmax}) is further cast to:
\begin{equation}
p\left( {y = k|{\bf x}} \right) = \frac{1}{{1 + \sum\limits_{j = 1,j \ne k}^K {\exp \left( {-h_{k,j} \left( {{\bf x},\Theta _D } \right)} \right)} }},
\label{eq_Dist}
\end{equation}
where
\begin{equation}
h_{k,j} \left( {{\bf x},\Theta _D } \right) = \phi _k \left( {{\bf x},\Theta _D } \right) - \phi _j \left( {{\bf x},\Theta _D } \right).
\label{eq_DistD}
\end{equation}

Comparing Eqs. (\ref{eq_Dist}), (\ref{eq_DistD}) and (\ref{eq_Gen2}), (\ref{eq_GenG}), we can see that $h_{k,j} \left( {{\bf x},\Theta _D } \right)$ can be connected to the $r_{k,j} \left( {{\bf x},\Theta _G } \right)$ with the LLR in calculation. This connection inspired us to incorporate the LLR of pairwise samples from a generative model to the neural network discriminative training for SV.
\subsection{Connecting generative and discriminative models through Log likelihood ratio for SV}
Based on the generative model, given a hypothesis $H_S$ or $H_D$, the joint probability of generating $({\bf x}_i ,{\bf x}_j$) is ${p\left( {{\bf x}_i ,{\bf x}_j |H_S } \right)}$ or ${p\left( {{\bf x}_i ,{\bf x}_j |H_D } \right)}$. In making a decision, the LLR is defined as:
 \begin{equation}
 r_{i,j} \mathop  = \limits^\Delta  r({\bf x}_i ,{\bf x}_j ) = \log \frac{{p({\bf x}_{i,} {\bf x}_j |H_S )}}{{p({\bf x}_{i,} {\bf x}_j |H_D )}}
 \label{eq_lllr}
 \end{equation}
 With a given decision threshold, we can decide whether the two observation vectors are from $H_S$ or $H_D$ (as defined in Eq. (\ref{eq_HSD})). For convenience of formulation, we define a trial as a tuple ${\bf{ z}}_{i,j}  = ( {{\bf x}_i ,{\bf x}_j } )$, and the two hypothesis spaces are constructed from the two data sets as:
\begin{equation}
\begin{array}{l}
 S = \left\{ {{\bf{z}}_{i,j}  = ( {{\bf x}_i ,{\bf x}_j } ) \in H_S } \right\} \\
 D = \left\{ {{\bf{z}}_{i,j}  = ( {{\bf x}_i ,{\bf x}_j } ) \in H_D } \right\} \\
 \end{array}
\label{eq_set}
\end{equation}
We first derive the LLR score calculation based on the JB based generative model.
\subsubsection{Joint Bayesian generative model approach}
Given an observation X-vector variable ${\bf x}$, it is supposed to be generated by a speaker identity variable and a random noise variable (possibly induced by different recording background noise, sessions, or transmission channels, etc.) as:
\begin{equation}
{\bf x} = {\bf u}  + {\bf n},
\label{eq_add}
\end{equation}
where $\bf u$ is a speaker identity vector variable, $\bf n$ represents intra-speaker variation caused by noise. For simplicity, the observation $\bf x$ is mean subtracted, and the speaker identity and intra-speaker variation variables are supposed to be with Gaussian distributions as:
\begin{equation}
\begin{array}{l}
 {\bf u} \sim N(0,{\bf C}_{\bf u} ) \\
 {\bf n} \sim N(0,{\bf C}_{\bf n} ) \\
 \end{array},
 \label{eq_gau}
\end{equation}
where ${\bf C}_{\bf u}$ and ${\bf C}_{\bf n}$ are speaker and noise covariance matrices, respectively. In verification, given a trial with ${\bf x}_i $ and ${\bf x}_j $ generated from Eq. (\ref{eq_add}), based on the assumption in Eq. (\ref{eq_gau}), the two terms ${p\left( {{\bf x}_i ,{\bf x}_j |H_S } \right)}$ and ${p\left( {{\bf x}_i ,{\bf x}_j |H_D } \right)}$ defined in Eq. (\ref{eq_lllr}) satisfy zero-mean Gaussian with covariances as:
\begin{equation}
\begin{array}{l}
 {\mathop{\rm cov}} _S  = \left[ {\begin{array}{*{20}c}
   {{\bf C}_{\bf u}  + {\bf C}_{\bf n} } & {{\bf C}_{\bf u} }  \\
   {{\bf C}_{\bf u} } & {{\bf C}_{\bf u}  + {\bf C}_{\bf n} }  \\
\end{array}} \right] \\
  \\
 {\mathop{\rm cov}} _D  = \left[ {\begin{array}{*{20}c}
   {{\bf C}_{\bf u}  + {\bf C}_{\bf n} } & 0  \\
   0 & {{\bf C}_{\bf u}  + {\bf C}_{\bf n} }  \\
\end{array}} \right] \\
 \end{array}
\end{equation}
Based on this formulation, the LLR defined in Eq. (\ref{eq_lllr}) could be calculated based on:
 \begin{equation}
 r( {{\bf x}_i ,{\bf x}_j } ) = {\bf x}_i^T {\bf Ax}_i  + {\bf x}_j^T {\bf Ax}_j  - 2{\bf x}_i^T {\bf Gx}_j,
 \label{eq_rij}
 \end{equation}
where
{\small
\begin{equation}
\begin{array}{l}
 {\bf A} \!=\! ({\bf C}_{\bf u} \! +\! {\bf C}_{\bf n} )^{ \!-\! 1} \! -\! [({\bf C}_{\bf u} \! +\! {\bf C}_{\bf n} ) \!-\! {\bf C}_{\bf u} ({\bf C}_{\bf u}  \!+\! {\bf C}_{\bf n} )^{\! - \!1} {\bf C}_{\bf u} ]^{\! -\! 1}  \\
 {\bf G} \!=\!  - (2{\bf C}_{\bf u}  \!+\! {\bf C}_{\bf n} )^{\! -\! 1} {\bf C}_{\bf u} {\bf C}_{\bf n}^{\! - \!1}  \\
 \end{array}
 \label{eq_jbAG}
\end{equation}
}
As seen from Eq. (\ref{eq_jbAG}), the generative model parameters ${\Theta _G }$ in estimating LLR are only related to covariance parameters ${\bf C}_{\bf u}$ and ${\bf C}_{\bf n}$ \cite{ChenJB2012,ChenJB2016}. Given a training data set, the parameters could be estimated using an EM (or EM-like) learning algorithm based on:
\begin{equation}
\Theta _G^*  = \arg \mathop {\min }\limits_{\Theta _G }  - \sum\limits_i {\log p({\bf X}_i |\Theta _G )}
\label{eq_EstPar}
\end{equation}
where $\Theta _G  = \left\{ {{\bf C}_{\bf u} ,{\bf C}_{\bf n} } \right\}$, ${\bf X}_i$ is a collection of samples for speaker $i$.
\subsubsection{Pairwise discriminative model approach}
The binary classification task defined in Eq. (\ref{eq_HSD}) can be solved based on discriminative neural network modeling as formulated in Eqs. (\ref{eq_softmax}) and (\ref{eq_Dist}). In neural network modeling, the parameters are neural weights (affine transform matrices with linear or nonlinear activations). We can connect the model parameters of a generative model with the neural weights and optimize them with an objective function.
As a binary classification task, given a trial with two observation X-vector variables ${\bf z}_{i,j}  = ({\bf x}_i {\rm ,}{\bf x}_j )$, the classification task is to estimate and compare $p(H_S |{\bf z}_{i,j} )$ and $p(H_D |{\bf z}_{i,j} )$. In the discriminative learning, the label is defined as:
\begin{equation}
y_{i,j}  = \left\{ {\begin{array}{*{20}c}
   { 1,{\bf z}_{i,j}  \in H_S }  \\
   { 0,{\bf z}_{i,j}  \in H_D }  \\
\end{array}} \right.
 \label{eq_H1H0}
\end{equation}
With reference to Eqs. (\ref{eq_Dist}) and (\ref{eq_DistD}), the posterior probability is estimated based on:
\begin{equation}
p(y_{i,j} |{\bf z}_{i,j} ) = \left\{ \begin{array}{l}
 \frac{1}{{1 + \exp ( - h_{H_S ,H_D } ({\bf z}_{i,j} ,\Theta _D ))}};{\bf z}_{i,j}  \in H_S  \\
 1 - \frac{1}{{1 + \exp ( - h_{H_S ,H_D } ({\bf z}_{i,j} ,\Theta _D ))}};{\bf z}_{i,j}  \in H_D  \\
 \end{array} \right.
\label{eq_DisB}
\end{equation}

As we have revealed from Eqs. (\ref{eq_Gen2}), (\ref{eq_GenG}), and (\ref{eq_lllr}), we replace the ${h_{H_S ,H_D } ({\bf z}_{i,j} ,\Theta _D )}$ with a LLR score, and define a mapping as a logistic function with scaled parameters as used in \cite{Platt1999,HLin2007}:
\begin{equation}
f\left( {r_{i,j} } \right)\mathop  = \limits^\Delta  \frac{1}{{1 + \exp \left( { - \left( {\alpha  r_{i,j}  + \beta } \right)} \right)}}
\label{eq_JointDG}
\end{equation}
where $r_{i,j}  = r({\bf z}_{i,j} )=   r({\bf x}_i ,{\bf x}_j )$ as defined in Eq. (\ref{eq_lllr}), and $\alpha$ and $\beta$ are gain and bias factors used in the regression model. In Eq. (\ref{eq_JointDG}), we integrated the LLR score estimated from the JB generative model in a discriminative training framework. The probability estimation in Eq. (\ref{eq_DisB}) is cast to:
\begin{equation}
\hat y_{i,j} \mathop  = \limits^\Delta  p(y_{i,j} |{\bf z}_{i,j} ) = \left\{ {\begin{array}{*{20}c}
   {f(r_{i,j} );{\bf z}_{i,j}  \in H_S }  \\
   {1 - f(r_{i,j} );{\bf z}_{i,j}  \in H_D }  \\
\end{array}} \right.
\end{equation}
The training can be based on optimizing the binary cross entropy defined as:
\begin{small}
\begin{equation}
L \!=\!  \!- \sum\limits_{{\bf z}_{i,j}  \in \{ H_S  \cup H_D \} }\! {(y_{i,j} \log f(r_{i,j} ) + (1 - y_{i,j} )\log (1 - f(r_{i,j} )))}
\label{eq_CrossEntropy}
\end{equation}
\end{small}
In the following subsection, we investigate the neural network architecture for the hybrid model framework.
\subsection{Coupling generative model parameters with neural network architecture}
\begin{figure}[tb]
\centering
\includegraphics[scale=1]{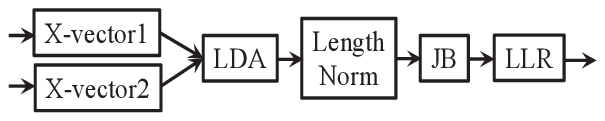}
\caption{Pipeline for joint Bayesian based generative modeling on X-vectors for speaker verification, LDA: Linear Discrimination Analysis, JB: Joint Bayesian model, LLR: log likelihood ratio.}
\label{figJB}
\vspace{2mm}
\end{figure}

The conventional state of the art framework for SV based on X-vector and JB model is illustrated in Fig. \ref{figJB}.
In this figure, the LDA is applied on the X-vector for discriminative feature extraction and dimension reduction. After the LDA, a vector length normalization is used, then a JB based generative model is applied by which the LLR is estimated. In a pairwise discriminative learning framework, the LLR can be used for a binary classification task, and could be implemented with a discriminative neural network model.

We first explain the LDA which will be approximated by an affine transform as used in the neural network modeling. For input X-vector samples and their corresponding labels $\left\{ {( {{\bf x}_1 ,{ y}_1 } ),( {{\bf x}_2 ,{y}_2 } ),...,( {{\bf x}_M ,{y}_M } )} \right\}$,  where ${\bf x}_i  \in {\rm R}^l$, and $M$ is the number of samples, the LDA transform is:
\begin{equation}
{\bf h}_i  = {\bf W}^T {\bf x}_i,
\label{eq_LDA}
\end{equation}
where ${\bf W} \in {\rm R}^{l \times d}$, $l$ and $d$ are the dimensions of the input X-vectors and the transformed feature vectors, respectively. ${\bf W}$ is estimated as follows:
\begin{equation}
{\bf W}^*  = \arg \mathop {\max }\limits_{\bf W} tr( {\frac{{{\bf W}^T {\bf S}_{\rm b} {\bf W}}}{{{\bf W}^T {\bf S}_{\rm w} {\bf W}}}} ),
\end{equation}
where $tr(.)$ denotes the matrix trace operator, ${\bf S}_{\rm w}$ and ${\bf S}_{\rm b}$ are the intra-class and inter-class covariance matrices. From Eq. (\ref{eq_LDA}), we can see that the LDA can be implemented as a linear dense layer in neural network modeling.

\begin{figure}[tb]
\centering
\includegraphics[scale=1]{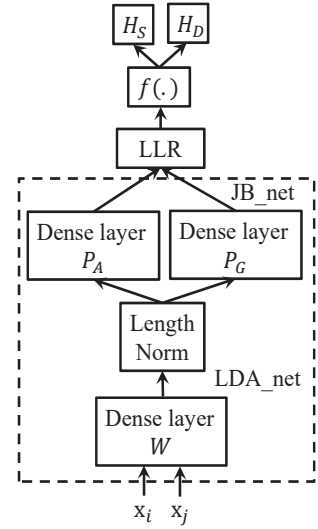}
\caption{The proposed two-branch Siamese neural network with coupling of the generative JB model structure for speaker verification (see the text for a detailed explanation). $H_D$: hypothesis for different speaker, $H_S$: hypothesis for the same speaker. Dense layers are with linear identity activation functions.}
\label{figNJBNet}
\end{figure}

We further look at the estimation of the LLR score defined in Eq. (\ref{eq_rij}). In Eq. (\ref{eq_rij}), $\bf A$ and $\bf G$ are negative semi-definite symmetric matrices \cite{ChenJB2012,ChenJB2016}, and they can be decomposed as:
\begin{equation}
\begin{array}{l}
 {\bf A} =  - {\bf P}_A {\bf P}_A^T  \\
 {\bf G} =  - {\bf P}_G {\bf P}_G^T  \\
 \label{eq_AGF}
 \end{array}
\end{equation}
The LLR score is cast to:
\begin{equation}
r_{i,j} = 2g_i^T g_j  - a_i^T a_i  - a_j^T a_j
\label{eq_Dlllr}
\end{equation}
with affine linear transforms as:
\begin{equation}
\begin{array}{l}
 a_i  = {\bf P}_A^T {\bf \tilde h}_i ,a_j  = {\bf P}_A^T {\bf \tilde h}_j  \\
 g_i  = {\bf P}_G^T {\bf \tilde h}_i ,g_j  = {\bf P}_G^T {\bf \tilde h}_j,  \\
 \label{eq_PAGF}
 \end{array}
\end{equation}
where the input to the JB model is the length normalized output from the LDA processing as:
 \begin{equation}
 \begin{array}{l}
 {\bf \tilde h}_i  = \frac{{{\bf h}_i }}{{\left\| {{\bf h}_i } \right\|}} \\
 {\bf \tilde h}_j  = \frac{{{\bf h}_j }}{{\left\| {{\bf h}_j } \right\|}} \\
 \end{array}
 \label{eq_lennorm}
 \end{equation}
The transforms in Eq. (\ref{eq_PAGF}) could be implemented in a neural network as linear dense layers. Based on these formulations, a two-branch Siamese neural network is designed as showed in Fig. \ref{figNJBNet}.
In this figure, there are two sub-nets, i.e., ``LDA\_net" and ``JB\_net". The ``LDA\_net" is a dense layer net with a transform $\bf W$ according to Eq. (\ref{eq_LDA}). In the ``JB\_net", the JB model structure is taken into consideration as two-branch $({\bf P}_A ,{\bf P}_G )$ dense layer network according to Eq. (\ref{eq_PAGF}). In training the Siamese neural network, the ``negative" and ``positive" samples are constructed as we did in pairwise discriminative training for language recognition task \cite{LUCLS2017}. In the generative model based backend, a length normalization block is often applied with the purpose of variable Gaussianization for the convenience of generative probability modeling \cite{lennorm}. In our proposed Siamese neural network backend model, the length normalization block is also used. The purpose is twofold: the first one is to exactly fit the generative model structure to the proposed discriminative neural network framework, the second one is to stabilize the neural network training by serving as a nonlinear transform for dynamic range normalization of neural activations.
\subsection{Direct evaluation metric (DEM): learning objective function based on minimum empirical Bayes risk (EBR)}
\begin{figure}[tb]
\vspace{4mm}
\centering
\includegraphics[scale=1]{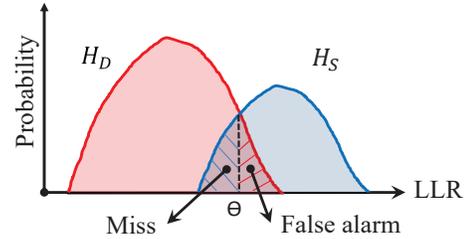}
\caption{The LLR distributions in $H_S$ and $H_D$ for the same and different speaker spaces, and two types of errors in the hypothesis test for SV.}
\label{figMissAlarm}
\end{figure}

\begin{figure}[tb]
\vspace{4mm}
\centering
\includegraphics[scale=1]{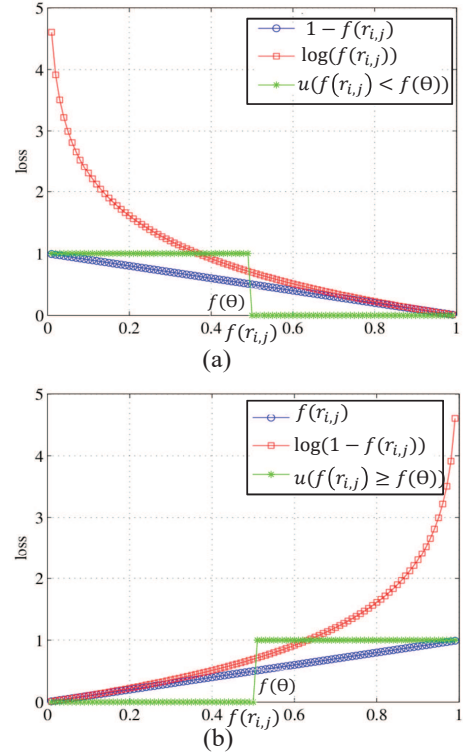}
\caption{Loss functions defined in Eqs. (\ref{eq_index}), (\ref{eq_lossscore})  and (\ref{eq_cesd}): (a) miss loss, and (b) false alarm loss\vspace{-0.2cm}}
\label{figMissAlarmLoss}
\end{figure}

The cross entropy defined in Eq. (\ref{eq_CrossEntropy}) can be applied for discriminative training in order to measure the classification error. However, the hypothesis test defined in Eq. (\ref{eq_HSD}) is different from a classification goal, and the final evaluation metric for SV usually adopts some different criterions. It is better to optimize model parameters directly based on the evaluation metrics. As a hypothesis test, there are two types of errors, i.e., type I and type II errors  \cite{BOSARIS2011, Lehmann2005}. The two types of errors are defined as:
\begin{equation}
\begin{array}{l}
 {\text{Type I error (false alarm):  }}{\bf z}_{i,j}  \in H_D ,{\text {LLR}}  \ge \theta  \\
 {\text{Type II error (miss):           }}{\bf z}_{i,j}  \in H_S ,{\text {LLR}}  < \theta,  \\
 \end{array}
 \label{eq_typeI_II}
\end{equation}
where $\theta$ is a decision threshold. These two types of errors are further illustrated in Fig. \ref{figMissAlarm} for a SV task.
In this figure, the objective for SV is to minimize the target miss $P_{{\rm miss}}$  (or false reject) and false alarm $P_{{\rm fa}}$ (or false accept) in the two hypothesis spaces $H_S$ and $H_D$. By selecting different decision thresholds, a detection error tradeoff (DET) graph could be obtained. In real applications, it is better to generalize the classification errors to a weighing of these two types of errors. With consideration of the prior knowledge in a measure of empirical Bayes risk (EBR), the evaluation metric for SV adopts a detection cost function (DCF) to measure the hardness of the decisions \cite{BOSARIS2011}.  It is defined as a weighted loss:
\begin{equation}
C_{\det }  \mathop  = \limits^\Delta  P_{{\rm tar}} C_{{\rm miss}} P_{{\rm miss}}  + ( {1 - P_{{\rm tar}} } )C_{{\rm fa}} P_{{\rm fa}},
\label{eq_cdet}
\end{equation}
where $C_{{\rm miss}}$ and $C_{{\rm fa}}$ are user assigned costs for miss and false alarm detections, $P_{{\rm tar}}$ is a prior of target trials, $P_{{\rm miss}}$ and $P_{{\rm fa}}$ are miss and false alarm probabilities defined as:
\begin{equation}
\begin{array}{l}
 P_{{\rm fa}}  = \frac{1}{{N_{{\rm non}} }}\sum\limits_{{\bf z}_{i,j}  \in H_D } {u(r_{i,j}  \ge \theta) }  \\
 P_{{\rm miss}}  = \frac{1}{{N_{{\rm tar}} }}\sum\limits_{{\bf z}_{i,j}  \in H_S } {u(r_{i,j}  < \theta) }  \\
 \end{array}
 \label{eq_index}
\end{equation}
In Eq. (\ref{eq_index}),  ``${N _{{\rm non}} }$" and ``${N _{{\rm tar}} }$" are the numbers of non-target and target trials, $ r_{i,j}$ is the LLR estimated from Eq. (\ref{eq_Dlllr}), $\theta$ is a decision threshold, and $u( . )$ is an indictor function for counting the number of trials with scores lower or higher than the decision threshold. In order to change the objective function to be differentiable and thus can be used in gradient based neural network learning, Eq. (\ref{eq_index}) is approximated by:
\begin{equation}
\begin{array}{l}
 P_{{\rm fa}}  \approx \frac{1}{{N_{{\rm non}} }}\sum\limits_{{\bf z}_{i,j}  \in \{ H_S  \cup H_D \} } {(1 - y_{i,j} )f(r_{i,j} )}  \\
 P_{{\rm miss}}  \approx \frac{1}{{N_{{\rm tar}} }}\sum\limits_{{\bf z}_{i,j}  \in \{ H_S  \cup H_D \} } {y_{i,j} (1 - f(r_{i,j} ))}  \\
 \end{array}
 \label{eq_lossscore}
\end{equation}
where $f(r_{i,j} )$ is a sigmoid logistic function defined the same as in Eq. (\ref{eq_JointDG}). With regard to the cross-entropy loss defined in Eq. (\ref{eq_CrossEntropy}), the weighted binary cross entropy loss (WBCE) can be formulated as:
\begin{equation}
L_{CE}  = w_S L_{CE\_H_S }  + (1 - w_S )L_{CE\_H_D },
\label{eq_wce}
\end{equation}
where $w_S$ is the weighting coefficient (as prior of target trials $P_{{\rm tar}}$ ), and the target and non-target cross entropy losses are defined as:
\begin{equation}
\begin{array}{l}
 L_{CE\_H_S }  =  - \frac{1}{{N_{{\rm tar}} }}\sum\limits_{{\bf z}_{i,j}  \in H_S } {y_{i,j} \log f(r_{i,j} )}  \\
 L_{CE\_H_D }  = \frac{1}{{N_{{\rm non}} }}\sum\limits_{{\bf z}_{i,j}  \in H_D } {(1 - y_{i,j} )\log (1 - f(r_{i,j} ))}  \\
 \end{array}
 \label{eq_cesd}
\end{equation}
For a further analysis, we show the loss functions defined in Eqs. (\ref{eq_index}), (\ref{eq_lossscore})  and (\ref{eq_cesd})  in Fig. \ref{figMissAlarmLoss}. From this figure, we can see that the losses defined in Eqs. (\ref{eq_lossscore})  and (\ref{eq_cesd}) have the same monotonic tendency in measuring the loss, and can be regarded as a soft loss of the miss and false alarm as defined in Eq. (\ref{eq_index}). In addition, from this figure, we can see that the calibrated LLR threshold $f(\theta )$ in loss score estimation is $0.5$. Based on this analysis, the definition of losses in Eq. (\ref{eq_cdet}) can be regarded as a generalized weighted loss from the definition of Eq. (\ref{eq_wce}).

In mini-batch based gradient back-propagation (BP) learning, the gradient is still estimated based on the chain rule from composition functions. For convenience of analysis, we reformulate the LLR score defined in Eq. (\ref{eq_rij}) as:
\begin{equation}
r_{i,j}  = {\bf \tilde h}_i^T {\bf A\tilde h}_i  + {\bf \tilde h}_j^T {\bf A\tilde h}_j  - 2{\bf \tilde h}_i^T {\bf G\tilde h}_j,
\label{eq_rij_re}
\end{equation}
where ${\bf \tilde h}_i$ and ${\bf \tilde h}_j$ are the length normalized vectors defined in Eq. (\ref{eq_lennorm}). Then the gradients for JB\_net parameters are derived as:
\begin{equation}
\begin{array}{l}
 \Delta {\bf P}_A \propto \frac{{\partial C_{{\rm det}} }}{{\partial f(r_{i,j} )}}\frac{{\partial f(r_{i,j} )}}{{\partial r_{i,j} }}\frac{{\partial r_{i,j} }}{{\partial {\bf A}}}\frac{{\partial {\bf A}}}{{\partial {\bf P}_A }} \\
 \Delta {\bf P}_G  \propto \frac{{\partial C_{{\rm det}} }}{{\partial f(r_{i,j} )}}\frac{{\partial f(r_{i,j} )}}{{\partial r_{i,j} }}\frac{{\partial r_{i,j} }}{{\partial {\bf G}}}\frac{{\partial {\bf G}}}{{\partial {\bf P}_G }} \\
 \end{array}
 \end{equation}
 And the gradients for LDA\_net parameters are derived as:
 \begin{equation}
 \Delta {\bf W} \propto \frac{{\partial C_{{\rm det}} }}{{\partial f(r_{i,j} )}}\frac{{\partial f(r_{i,j} )}}{{\partial r_{i,j} }}\left( {\frac{{\partial r_{i,j} }}{{\partial {\bf \tilde h}_i }}\frac{{\partial {\bf \tilde h}_i }}{{\partial {\bf h}_i }}\frac{{\partial {\bf h}_i }}{{\partial {\bf W}}} + \frac{{\partial r_{i,j} }}{{\partial {\bf \tilde h}_j }}\frac{{\partial {\bf \tilde h}_j }}{{\partial {\bf h}_j }}\frac{{\partial {\bf h}_j }}{{\partial {\bf W}}}} \right)
 \end{equation}
Following the definition function in each term, the gradients can be deduced which are used in BP based neural network learning.
\section{Experiments and results}
\label{sec-III}
We carried out experiments on SV tasks where the test data sets are from Speakers in the wild (SITW) \cite{SITW2016} and Voxceleb \cite{Vox2020}. The speaker features and models were trained on Voxceleb (sets 1 and 2) \cite{Vox2020}. A state of the art pipeline for constructing the SV system is adopted as shown in Fig. \ref{figJB}. In this figure, the ``LDA", ``Length Norm", and ``JB" blocks are designed independently rather than optimized jointly. The input speaker features in our pipeline are X-vectors. The X-vectors are extracted based on a well trained neural network model which is designed for a speaker classification task \cite{Snyder2018}. As backend models, both the well-known PLDA and JB based generative models are implemented in our comparisons.
\subsection{Speaker embedding features based on X-vectors}
\label{exp_x-vector}
A speaker embedding model is trained for the X-vector extraction. The neural network architecture of the embedding model is composed of deep time delay neural network (TDNN) layers and statistical pooling layers implemented the same as introduced in \cite{Snyder2018} and used in Kaldi \cite{KaldiSV}. In training the model, the cross entropy criterion for speaker classification is used as the learning objective function. The training data set includes two data sets from Voxceleb corpus, i.e., the training set of Voxceleb1 corpus from which speakers included in the test set of the SITW are removed, and the training set of Voxceleb2. In total, there are about 7,185 speakers with 1,236,567 utterances used for training. Moreover, data augmentation is applied by adding noise, music, babble with several SNRs, and reverberation with simulated room impulse response is also applied to increase data diversity \cite{KaldiSV}. Input features for training the speaker embedding model are MFCCs with 30 Mel band bins. The MFCCs are extracted with 25 ms frame length and 10 ms frame shift. Energy based voice activity detection (VAD) is applied to remove silence background regions in speaker feature extraction. More details of feature and model architecture and training procedures were introduced in \cite{Snyder2018}. The final extracted X-vectors are with 512 dimensions.
\subsection{Backend models}
\label{exp_backend}
Although X-vectors extracted from the speaker embedding model are supposed to encode speaker discriminative information, they also encode other acoustic factors. In a conventional pipeline as illustrated in Fig. \ref{figJB}, LDA is applied before applying a generative speaker model. In this study, the 512-dimension X-vectors are transformed to 200-dimension vectors by LDA. Correspondingly, in the discriminative neural network model as showed in Fig. \ref{figNJBNet}, a dense layer with 200 neurons is also applied. Moreover, in the discriminative model, two dense layers corresponding to ${\bf P}_A$ and ${\bf P}_G$  of the JB model are trained with ``positive" and ``negative" X-vector pairs (pairs from the same and different speakers). Since the discriminative neural network architecture fits well to the pipeline based on a generative model structure, the dense layer parameters could be initialized with the LDA and JB model parameters in training (according to Eqs. (\ref{eq_LDA}) and (\ref{eq_PAGF})). By this initialization, the discriminative training starts exactly from the model parameters of backend pipeline including the LDA and the generative model, and further refines the discriminability for the SV task. For comparison, the random initialization method with ``he\_normal" as widely used in deep neural network learning is also applied in experiments \cite{He2015}.  In model training, the Adam algorithm with an initial learning rate of $0.0005$ \cite{Adam} was used. In order to include enough ``negative" and ``positive" samples, the mini-batch size was set to 4096. The training X-vectors were splitted to training and validation sets with a ratio of $9:1$. The model parameters were selected based on the best performance on the validation set.
\subsection{Results}
We first carried out SV experiments on the data sets of SITW. Two test sets are used, i.e., development and evaluation sets, and each is used as an independent test set. The evaluation metrics are equal error rate (EER) and minimum decision cost function (minDCF) (with target priors 0.01 and 0.001) \cite{SITW2016}. The EER denotes when the type I and type II errors (as defined in Eq. (\ref{eq_typeI_II})) are equal, and the minDCF is optimized from the DCF defined in Eq. (\ref{eq_cdet}) (with $C_{{\rm miss}}=C_{{\rm fa}}=1$). In the optimization of the minDCF, the decision threshold for false alarm and miss error estimations is also jointly optimized. The performance results are showed in Tables \ref{tab1} and \ref{tab2}.
\begin{table}[hpbt]
\centering
\caption{Performance on the development set of SITW.}
\begin{tabular}{|c||c|c|c|}
\hline
 Methods&EER (\%)&minDCF (0.01)&minDCF (0.001)\\
\hline
\hline
LDA(200)+PLDA  &3.003&0.3315&0.5198\\
\hline
LDA(200)+JB  &3.043&0.3288&0.5019\\
\hline
Hybrid (rand init)  &4.159&0.3792&0.5883\\
\hline
Hybrid (JB init) &\textbf{2.662}&\textbf{0.2972}&\textbf{0.4466}\\
\hline
\end{tabular}
\label{tab1}
\end{table}\vspace{-0.2cm}
\begin{table}[hpbt]
\centering
\caption{Performance on evaluation set of SITW. }
\begin{tabular}{|c||c|c|c|}
\hline
 Methods&EER (\%)&minDCF (0.01)&minDCF (0.001)\\
\hline
\hline
LDA(200)+PLDA  &3.554&0.3526&0.5657\\
\hline
LDA(200)+JB  &3.496&0.3422&0.5645\\
\hline
Hybrid (rand init)  &4.505&0.3920&0.6003\\
\hline
Hybrid (JB init)  &\textbf{3.142}&\textbf{0.3075}&\textbf{0.4619}\\
\hline
\end{tabular}
\label{tab2}
\end{table}
\begin{table}[hpbt]
\centering
\caption{Performance on evaluation set of voxceleb1 test. }
\begin{tabular}{|c||c|c|c|}
\hline
 Methods&EER (\%)&minDCF (0.01)&minDCF (0.001)\\
\hline
\hline
LDA(200)+PLDA &3.128&0.3258&0.5003\\
\hline
LDA(200)+JB  &3.105&0.3226&0.4992\\
\hline
Hybrid (rand init)  &3.340&0.3778&0.4977\\
\hline
Hybrid (JB init) &\textbf{2.837}&\textbf{0.3011}&\textbf{0.3743}\\
\hline
\end{tabular}
\label{tab_voxceleb1}
\end{table}
In these two tables, ``LDA(200)+PLDA" and ``LDA(200)+JB" represent the PLDA and JB generative model based SV systems following the pipeline in Fig. \ref{figJB} (the dimension for LDA is 200) (replace the block ``JB" with ``PLDA" for the PLDA based SV system). ``Hybrid" denotes the proposed Siamese neural network backend based SV system which takes the JB model structure and parameter coupling in designing the neural model architecture following the pipeline in Fig. \ref{figNJBNet}. And in the ``Hybrid" SV system, two model initialization methods are tested in model training as explained in section \ref{exp_backend}. From these two tables, we can see that the performance of the JB based generative model is comparable or slightly better than that of the PLDA based model. In the hybrid model, if model parameters (``LDA\_net" and ``JB\_net") are randomly initialized, the performance is worse than the original generative model based results. However, when the neural network parameters are initialized with the ``LDA" and ``JB" based model parameters, the performance is significantly improved. These results indicate that the discriminative training could further enhance the discriminative power of the generative model when the model parameters are initialized with the generative model based parameters. Otherwise, random initialization in the discriminative learning does not enhance the performance even when the generative model structure is taken into consideration. Following the same process, the experimental results on voxceleb1 test set are showed in Table \ref{tab_voxceleb1}. From this table, we could observe the same tendency as in Tables \ref{tab1} and \ref{tab2}.
\subsection{Ablation study}
Many factors may contribute to the final performance in the proposed framework. In this paper, we consider two aspects which are directly related to our contributions: one is the hybrid generative and discriminative model architecture design, the other is the optimization objective function design. In the model architecture design, there are two important modeling blocks, i.e., ``LDA\_net" and ``JB\_net" as illustrated in Fig. \ref{figNJBNet}. The function of the ``LDA\_net" is extracting low-dimension discriminative speaker representations from X-vectors, whereas  the ``JB\_net" is applied on the extracted feature vectors for speaker modeling. They were jointly learned in a unified framework. In the optimization objective function design, although the direct evaluation metric could be regarded as a generalization from the weighted binary cross entropy function, the degree of penalty for miss and false alarm errors is different. In this subsection, we investigate their effects on SV performance one by one with ablation studies.
\subsubsection{Effect of the ``LDA\_net" in learning}
 X-vectors are extracted from a TDNN based speaker embedding model which is optimized for speaker classification. After the LDA process, the speaker feature has a strong power for speaker discrimination. In the proposed hybrid model, the LDA model is further jointly optimized for the SV task. We verify the discrimination power of speaker representations on SV performance with random setting of the ``JB\_net" while only setting the parameters of the ``LDA\_net" with the following conditions (after setting, the model is not further trained any more): (a) setting the ``LDA\_net" with the LDA parameters (independent LDA transform), (b) setting the ``LDA\_net" with the jointly trained LDA parameters. The results are showed in table \ref{tab_init}. From these results, we can see that after joint training (in setting (b)), the performance is further improved.
\begin{table}[hpbt]
\centering
\caption{Performance on the development set of SITW: random setting of the classifier model (``JB\_net") and with two setting conditions for the ``LDA\_net". Setting (a): independent LDA transform, Setting (b): jointly trained LDA transform.}
\begin{tabular}{|c||c|c|c|}
\hline
 LDA\_net setting&EER (\%)&minDCF (0.01)&minDCF (0.001)\\
\hline
\hline
Setting (a)  &8.24&0.6546&0.8434\\
\hline
Setting (b)  &\textbf{8.01}&\textbf{0.5968}&\textbf{0.7820}\\
\hline
\end{tabular}
\label{tab_init}
\end{table}
\subsubsection{Effect of $\bf A$ and $\bf G$ on SV performance}
\label{sect_AG_effect}
As showed in Eq. (\ref{eq_rij}), the two terms have different effects on the speaker verification performance. In our discriminative training which integrates the LLR of the JB model, the LLR in Eq. (\ref{eq_rij}) is adapted. With different settings of $\bf A$ and $\bf G$ on Eq. (\ref{eq_rij}), we could obtain:
\begin{equation}
r({\bf x}_i ,{\bf x}_j ) = \left\{ {\begin{array}{*{20}c}
   { - 2{\bf x}_i^T {\bf Gx}_j ;\text{ for }{\bf A} = 0}  \\
   {{\bf x}_i^T {\bf Ax}_i  + {\bf x}_j^T {\bf Ax}_j ;\text{ for }{\bf G} = 0}  \\
   {({\bf x}_i  - {\bf x}_j )^T {\bf G}({\bf x}_i  - {\bf x}_j );\text{ for }{\bf A} = {\bf G}}  \\
   {({\bf x}_i  - {\bf x}_j )^T {\bf A}({\bf x}_i  - {\bf x}_j );\text{ for }{\bf G} = {\bf A}}  \\
\end{array}} \right.
\label{eq_maha_convert}
\end{equation}
Based on this formulation, we could check the different effects of $\bf A$ and $\bf G$ on the SV performance. The two matrices $\bf A$ and $\bf G$  are connected to the two dense layer branches of the hybrid model with weights ${\bf P}_A$ and ${\bf P}_G$ (refer to Fig. \ref{figNJBNet}). In our model, the dense layers were first initialized with the parameters from the learned JB based generative model, then the model was further trained with pairwise ``negative" and ``positive" samples. Only in testing stage, we use different parameter settings for experiments according to Eq. (\ref{eq_maha_convert}), and the results are showed in Tables \ref{tab4} and \ref{tab5} for the dev set of SITW before and after discriminative training, respectively.
\begin{table}[hpbt]
\centering
\caption{Performance on the development set of SITW \textbf{before joint training}: setting the LDA\_net and JB\_net with the independently learned LDA and JB model parameters, with different experimental settings of classifier model (``JB\_net"). ).}
\begin{tabular}{|c||c|c|c|}
\hline
 Methods&EER (\%)&minDCF (0.01)&minDCF (0.001)\\
\hline
\hline
A (G=0)  &47.71&1.000&1.000\\
\hline
G (A=0) &6.353&0.8261&0.9806\\
\hline
A, G (set G to A) &\textbf{3.119}&\textbf{0.3604}&\textbf{0.5844}\\
\hline
A, G (set A to G) &3.504&0.3978&0.6316\\
\hline
\end{tabular}
\label{tab4}
\end{table}
\begin{table}[hpbt]
\centering
\caption{Performance on the development set of SITW \textbf{after joint training}: different experimental settings of classifier model (``JB\_net").}
\begin{tabular}{|c||c|c|c|}
\hline
 Methods&EER (\%)&minDCF (0.01)&minDCF (0.001)\\
\hline
\hline
A (G=0)  &50.29&0.9996&0.9996\\
\hline
G (A=0)  &4.775&0.4206&0.6340\\
\hline
A, G (set G to A)  &\textbf{2.811}&\textbf{0.2975}&0.4561\\
\hline
A, G (set A to G)  &3.080&0.3134&\textbf{0.4505}\\
\hline
\end{tabular}\vspace{-0.2cm}
\label{tab5}
\end{table}
In these two tables, by comparing conditions with ${\bf A}=0$ or ${\bf G}=0$, we can see that the cross term contributes more to the SV performance, i.e., the dense layer branch with neural weight ${\bf P}_G$ contributes the most discriminative information in the SV task. Moreover, when keeping the cross term either by setting ${\bf A} = {\bf G}$ or ${\bf G} = {\bf A}$, the performance is better than setting any one of them to be zero. In summary, the contribution of discriminative information from feature self-norm associated with matrix ${\bf A}$ is less while feature cross-term associated with ${\bf G}$ contributes most in the SV task.
\subsubsection{Relation to distance metric learning}
\begin{figure}[tb]
\centering
\includegraphics[scale=1]{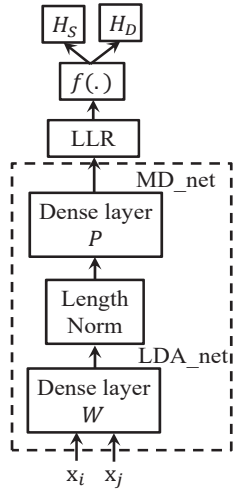}
\caption{Siamese neural network with Mahalanobis distance net (MD\_net) on X-vector features for speaker verification. Dense layers are with linear identity activation functions.\vspace{-0.2cm}}
\label{figNJBNet_one}
\end{figure}

Distance metric learning is widely used in discriminative learning with pairwise training samples as input \cite{Xing2002,Weinberger2009}. The Mahalanobis distance metric between two vectors is defined as:
\begin{equation}
d_{i,j} \mathop  = \limits^\Delta  d({\bf x}_i ,{\bf x}_j ) = ({\bf x}_i  - {\bf x}_j )^T {\bf M}({\bf x}_i  - {\bf x}_j ),
\label{eq_maha}
\end{equation}
where ${\bf M} = {\bf PP}^T$ is a positive definite matrix. Based on this distance metric, the binary classification task for SV can be formulated as:
\begin{equation}
p(y_{i,j} |{\bf z}_{i,j} ) = \sigma (\lambda (d_0  - d_{i,j} )),
\label{eq_mhad}
\end{equation}
where $\sigma (x) = (1 + \exp ( - x))^{ - 1}$ is the sigmoid logistic function, $d_0$ is a distance decision threshold, and $\lambda$ is a scale parameter for probability calibration. From Eq. (\ref{eq_mhad}), we can see that when the Mahalanobis distance $d({\bf x}_i ,{\bf x}_j ) <d_0$, the probability of ${\bf x}_i$ and ${\bf x}_j$ belonging to the same speaker is high, and vice versa. With pairwise ``positive" and ``negative" samples, the parameters ($\bf M$, $d_0$, and $\lambda$) can be learned based on a given training data set as a binary discriminative learning task. Comparing Eqs. (\ref{eq_maha}) and (\ref{eq_maha_convert}), we can see that if we set ${\bf A}={\bf G}$ or ${\bf G}={\bf A}$ , the LLR and Mahalanobis distance have the same formulation form (except the difference in matrix as negative or positive definite), i.e., $d({\bf x}_i ,{\bf x}_j ) \propto  - r({\bf x}_i ,{\bf x}_j )$. In this sense, the distance metric based discriminative learning framework can be regarded as a special case of the hybrid discriminative framework, and the LLR defined in Eq. (\ref{eq_lllr}) is cast to:
\begin{equation}
r({\bf x}_i ,{\bf x}_j ) = \log \frac{{p(\Delta _{i,j} |H_S )}}{{p(\Delta _{i,j} |H_D )}},
\label{eq_bayesian}
\end{equation}
where $\Delta _{i,j}  = {\bf x}_i  - {\bf x}_j$. From this definition, we can see that the distance metric based discriminative learning only considers the distribution of the pairwise sample distance space \cite{BayesianFace}. In our implementation, by merging the two dense layers of the classifier model (``JB\_net" with parameters ${\bf P}_A$ and ${\bf P}_G$), the proposed hybrid framework is changed to be one branch framework as showed in Fig. \ref{figNJBNet_one}.
In this figure, the ``MD\_net" is the network dense layer for Mahalanobis distance metric with an affine transform matrix $\bf P$, and it can be initialized with the parameters of the JB based generative model (either ${\bf P} = {\bf P}_A$ or ${\bf P} = {\bf P}_G$), or with random values (he\_normal). We test this one branch model on the dev set of SITW with different settings of the ``MD\_net" (the ``LDA\_net" is initialized with the LDA transform based parameters), and show the results in Table \ref{tab_dm}.
\begin{table}[hpbt]
\centering
\caption{Performance on the development set of SITW based on the Siamese neural network with ``MD\_net" as classifier model. }
\begin{tabular}{|c||c|c|c|}
\hline
 Methods&EER (\%)&minDCF (0.01)&minDCF (0.001)\\
\hline
\hline
Random init $\bf P$ &3.966&0.3743&0.5543\\
\hline
Init ${\bf P}$ with ${\bf P}_A$  &\textbf{3.621}&\textbf{0.3686}&\textbf{0.5472}\\
\hline
Init ${\bf P}$ with ${\bf P}_G$  &4.005&0.4060&0.6003\\
\hline
\end{tabular}
\label{tab_dm}
\end{table}
From this table, we can see that when the LDA\_net and MD\_net of the one branch model are initialized with the LDA and ${\bf P}_A$ parameters, the performance is the best. However, no matter in what conditions, comparing results in Tables \ref{tab1} and \ref{tab_dm}, we can see that the hybrid model framework showed the best performance which confirmed that the model structure inspired by the JB based generative model is helpful in the SV task.
\begin{figure}[tb]
\centering
\includegraphics[scale=1]{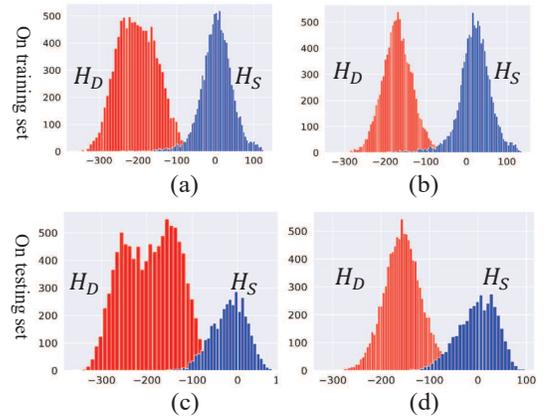}
\caption{LLR distributions in $H_S$ and $H_D$ spaces: the first row (a, b) for the training set, the second row (c, d) for the testing set; the left column (a and c) for model setting with generative model parameters learned based on the EM algorithm, and the right column (b and d) for model setting with discriminatively trained parameters after initializing with generative model parameters learned based on the EM algorithm.\vspace{-0.2cm}}
\label{fig_H0H1_AG_train_eval}
\end{figure}
\subsubsection{LLR distributions for intra- and inter-speaker spaces}
As defined in Eq. (\ref{eq_lllr}), the performance is measured based on the LLR distributions in two spaces, i.e., the intra-speaker space $H_S$ and inter-speaker space $H_D$. The separability can be visualized as the histogram distributions of pairwise distances in the two spaces. We check the histograms of the LLR on the training and test sets based on the hybrid model (refer to network pipeline in Fig. \ref{figNJBNet}) with different parameter settings, and show them in Fig. \ref{fig_H0H1_AG_train_eval}. From this figure, we can see that with the discriminative training, the separation is further enhanced. In particular, the LLR distribution of ``negative" sample pairs becomes much more compact for both training and testing data sets.
\subsubsection{Effect of objective function design}
Different objective function may affect the optimization process and hence may result in different performance. Although the direct evaluation metric (DEM) defined in Eqs. (\ref{eq_cdet}), (\ref{eq_lossscore}) can be regarded as a generalization of the weighted binary WBCE (defined in Eqs. (\ref{eq_wce}), (\ref{eq_cesd})), theoretically, the performance based on optimizing this DEM should be better than based on optimizing the WBCE due to the measurement consistency in both training and testing. We carried out experiments to test model performance when the model was optimized with DEM and WBCE based objective functions, and show the results in Tables \ref{tabBCEDEV} and \ref{tabBCEEVAL} for the development and evaluation sets of SITW, respectively.
\begin{table}[hpbt]
\centering
\caption{Performance with different optimization objective functions (on the development set of SITW).}
\begin{tabular}{|c||c|c|c|}
\hline
 Objective functions&EER (\%)&minDCF (0.01)&minDCF (0.001)\\
\hline
\hline
WBCE (Eqs. \ref{eq_wce}, \ref{eq_cesd})   &2.695&0.3157&0.5670\\
\hline
DEM (Eqs. \ref{eq_cdet}, \ref{eq_lossscore})&\textbf{2.662}&\textbf{0.2972}&\textbf{0.4466}\\
\hline
\end{tabular}
\label{tabBCEDEV}
\end{table}
\begin{table}[hpbt]
\centering
\caption{Performance with different optimization objective functions (on the evaluation set of SITW).}
\begin{tabular}{|c||c|c|c|}
\hline
 Objective functions&EER (\%)&minDCF (0.01)&minDCF (0.001)\\
\hline
\hline
WBCE (Eqs. \ref{eq_wce}, \ref{eq_cesd})   &\textbf{3.089}&0.3567&0.6163\\
\hline
DEM  (Eqs. \ref{eq_cdet}, \ref{eq_lossscore}) &3.142&\textbf{0.3075}&\textbf{0.4619}\\
\hline
\end{tabular}\vspace{-0.2cm}
\label{tabBCEEVAL}
\end{table}
In these two tables, the model parameters are initialized from the JB based generative model, and re-trained based on the two objective functions (setting prior of target trials to $0.01$). From these two tables, we can confirm that the direct evaluation metric is much more suitable in discriminative training for the SV tasks especially in terms of minDCF.\vspace{-0.2cm}
\section{Discussion}
\label{sec-IV}
As a detection task, the performance of SV could be benefitted from many aspects. For example, effective neural network architectures for X-vector extraction \cite{ResNetSV1,ResNetSV2,ResNetSV3,ResNetSV4}, advanced data augmentation for speaker classification training in robust X-vector extraction \cite{Snyder2018, DAUSV1}, borrowing the idea of better margin based objective functions from \cite{NewSV1} for training speaker embedding models \cite{NewSV2,NewSV3}. Particularly, integrating attention models with the most advanced techniques for X-vector extraction shows significant improvement in SV tasks \cite{ECAPA}. The improvement is largely due to the effective exploration of the speaker discriminative information in modeling and learning. Since our proposed discriminative learning framework in this paper is also for enhancing the discriminative power for SV, it is natural to wander: whether the proposed hybrid backend is still effective or not when strong X-vector features are used as inputs. We carried out additional experiments by using X-vectors extracted from ECAPA-TDNN \cite{ECAPA} as inputs to our proposed framework. The ECAPA-TDNN was trained using Voxceleb (training sets 1 and 2).
\begin{table}[hpbt]
\centering
\caption{Performance on the evaluation set of voxceleb1 test (X-vectors extracted from ECAPA-TDNN \cite{ECAPA}). }
\begin{tabular}{|c||c|c|c|}
\hline
 Methods&EER (\%)&minDCF (0.01)&minDCF (0.001)\\
\hline
\hline
LDA (150)+JB  &1.280&0.2013&0.3287\\
\hline
LDA (180)+JB  &1.186&0.1844&0.3055\\
\hline
LDA (192)+JB  &1.170&0.1858&0.2846\\
\hline
\hline
Hybrid (JB init) &\textbf{0.803}&\textbf{0.0919}&\textbf{0.1249}\\
\hline
\end{tabular}\vspace{-0.2cm}
\label{tab_voxceleb1_ecapa}
\end{table}
Different from the X-vectors extracted in \cite{Snyder2018} (with dimension $512$), the dimension of X-vectors extracted from ECAPA-TDNN is $192$. Before designing the hybrid neural backend framework, we first examined the effective dimensions as used in LDA. The results are showed in Table \ref{tab_voxceleb1_ecapa} (as ``LDA (dimension)+JB"). From this table, we can see that using full dimensions achieved the best performance. Therefore, in designing the Siamese neural network for backend modeling according to Fig. \ref{figNJBNet}, the dimensions for LDA\_net and JB\_net were set to 192 neural nodes. All other settings were kept the same as used in experiments in Section \ref{sec-III}. The results are showed as ``Hybrid (JB init)" in Table \ref{tab_voxceleb1_ecapa}. From these results, we can see that there is a large improvement by using the ECAPA-TDNN based X-vector extraction system, and the proposed neural network based backend still provided significant improvement on this strong baseline system. Our results were better or comparable to the best settings in  \cite{ECAPA} (EER (\%)= 0.87, minDCF (0.01)=0.1066) (please note that the settings in training the X-vector model, especially the backend pipelines were different).\vspace{-0.2cm}
\section{Conclusion and future work}
\label{sec-conclud}
The current state of the art pipeline for SV is composed of two building models, i.e., a front-end model for speaker feature extraction, and a generative model based backend model for speaker verification. In this study, the X-vector as a speaker embedding feature is extracted in the front-end model which encodes strong speaker discriminative information. Based on this speaker feature, a JB based generative backend model is applied. The JB model tries to model the probability distributions of speaker features, and could predict the conditional probabilities for utterances even from unknown speakers. But as a generative model, the parameter estimation can be easily distracted with  nuisance features in a high dimensional space. As an alternative, the SV task can be also regarded as a binary classification task. Correspondingly, a discriminative learning framework can be applied with ``positive" and ``negative" sample pairs (as from the same speaker and different speakers). Under a discriminative learning framework, discriminative features can be automatically transformed and modeled in a unified optimization framework. In this study, as our main contribution, we proposed to couple the generative model structure and parameters with the dense layers of a neural network learning framework as a hybrid model. The key point is that we reformulated the LLR estimation in the JB model to a distance metric as used in the discriminative learning framework. In particular, the linear matrices in the JB model are factorized to be the linear affine transforms as implemented in dense layers of the neural network model. And the network parameters are connected to the JB model parameters so that they could be initialized by the generatively learned parameters. Moreover, as our second contribution to the discriminative learning framework, rather than simply learning the hybrid model with a conventional binary discrimination objective function, the direct evaluation metric for hypothesis test with consideration of false alarm and miss errors was applied as an objective function in parameter optimization learning.

In this study, the JB based generative model is based on simple Gaussian probability distribution assumptions of speaker features and noise. In real applications, the probability distributions are much more complex. Although it is difficult for a generative model to fit complex shapes of probability distributions in a high dimensional space, it is relatively easy for a discriminative learning framework to approximate the complex distribution shapes. In the future, we will extend the current study for a hybrid model framework to learn more complex probability distributions in SV tasks.

\section{Acknowledgment}
Authors would like to thank the anonymous reviewers for helping us in paper revision. The work is partially supported by JSPS KAKENHI No. 19K12035, No. 21K17776.\vspace{-0.2cm}

\bibliographystyle{IEEEbib}
\bibliography{strings,refs}

\begin{thebibliography}{1}
\bibitem{Hansen2015}
J. Hansen, T. Hasan, ``Speaker recognition by machines and humans: A tutorial review," \emph{IEEE Signal processing magazine}, vol. 32, no. 6, pp. 74-99, 2015.

\bibitem{Beigi2007}
H. Beigi, Fundamentals of Speaker Recognition, Springer-Verlag, Berlin, 2011, ISBN 978-0-387-77591-3.

\bibitem{Dehak2011}
N. Dehak, P. Kenny, R. Dehak, P. Dumouchel, and P. Ouellet, ``Front-end factor analysis for speaker verification," \emph{IEEE Transactions on Audio, Speech, and Language Processing}, vol. 19, no. 4, pp. 788-798, 2011.

\bibitem{Variani2014}
E. Variani, X. Lei, E. McDermott, I. L. Moreno and J. Gonzalez-Dominguez, ``Deep neural networks for small footprint text-dependent speaker verification," \emph{IEEE International Conference on Acoustics, Speech and Signal Processing (ICASSP)},  pp. 4052-4056, 2014.

\bibitem{Snyder2018}
D. Snyder, D. Garcia-Romero, G. Sell, D. Povey, and S. Khudanpur, ``X-vectors: Robust DNN embeddings for speaker recognition," in \emph{IEEE International Conference on Acoustics, Speech and Signal Processing (ICASSP)}, pp. 5329-5333, 2018.

\bibitem{SugiyamaFLDA}
M. Sugiyama, "Local Fisher discriminant analysis for supervised dimensionality reduction," in \emph{Proc. of the 23rd international conference on Machine learning (ICML)}, pp. 905-912, 2006.

\bibitem{ShenICASSP2016}
P. Shen, X. Lu, L. Liu, H. Kawai, ``Local fisher discriminant analysis for spoken language identification," in \emph{Proc. of ICASSP}, pp. 5825-5829, 2016.

\bibitem{Prince2007}
S. Prince and J. Elder, ``Probabilistic linear discriminant analysis for inferences about identity," in \emph{IEEE International Conference on Computer Vision (ICCV)}, pp. 1-8, 2007.

\bibitem{PLDAUni}
A. Sizov, K. Lee, and T. Kinnunen, ``Unifying probabilistic linear discriminant analysis variants in biometric authentication," in Joint IAPR International Workshops on Statistical
Techniques in Pattern Recognition (SPR) and Structural and Syntactic Pattern Recognition (SSPR). Springer, pp. 464-475, 2014.

\bibitem{Kenny2010}
P. Kenny, ``Bayesian speaker verification with heavy tailed priors," in Odyssey Speaker and Language Recognition Workshop: 14, 2010.

\bibitem{Kenny2013}
P. Kenny, T. Stafylakis, P. Ouellet, M. J. Alam and P. Dumouchel, ``PLDA for speaker verification with utterances of arbitrary duration," in \emph{Proc. of  ICASSP}, pp. 7649-7653, 2013.

\bibitem{ChenJB2012}
D. Chen, X. Cao, L. Wang, F. Wen, and J. Sun, ``Bayesian face revisited: A joint formulation," in \emph{European Conference on Computer Vision}, pp.
566-579, 2012.

\bibitem{ChenJB2016}
D. Chen, X. Cao, D. Wipf, F. Wen, and J. Sun, ``An efficient joint formulation for Bayesian face verification," \emph{IEEE Transactions on pattern analysis and machine intelligence}, vol. 39, pp. 32-46, 2016.

\bibitem{WangOu2017}
Yiyan Wang, Haotian Xu, Zhijian Ou, "Joint Bayesian Gaussian Discriminant Analysis for speaker verification," in \emph{Proceeding of ICASSP}, pp. 5390-5394, 2017.

\bibitem{Vox2020Web}
A. Nagrani, J. Chung, J. Huh, A. Brown, E. Coto, W. Xie, M. McLaren, D. Reynolds and A. Zisserman, "VoxSRC 2020: The Second VoxCeleb Speaker Recognition Challenge," https://www.robots.ox.ac.uk/~vgg/data/voxceleb/competition2020.html.

\bibitem{WANSVM2000}
V. Wan, W. Campbell, ``Support vector machines for speaker verification and identification," Neural Networks for Signal Processing X, in \emph{Proceedings of the IEEE Signal Processing Society Workshop}, vol. 2, pp. 775-784, 2000.

\bibitem{SVMSV3}
W. Campbell, D. Sturim, and D. Reynolds, ``Support vector machines using GMM supervectors for speaker verification," IEEE signal processing letters, vol. 13, no. 5, pp. 308-311, 2006.

\bibitem{Villalba2017}
J. Villalba, N. Brummer, N. Dehak, ``Tied variational autoencoder backends for i-vector speaker recognition," in \emph{Proc of INTERSPEECH}, pp. 1004-1008, 2017.

\bibitem{Burget2011}
L. Burget, O. Plchot, S. Cumani, O. Glembek, P. Matejka and N. Brummer, ``Discriminatively trained Probabilistic Linear Discriminant Analysis for speaker verification," in \emph{Proceeding of IEEE International Conference on Acoustics, Speech and Signal Processing (ICASSP)}, pp. 4832-4835, 2011.

\bibitem{Cumanni2013}
S. Cumani, N. Brummer, L. Burget, P. Laface, O. Plchot and V. Vasilakakis, ``Pairwise Discriminative Speaker Verification in the I-Vector Space," \emph{IEEE Transactions on Audio, Speech, and Language Processing}, vol. 21, no. 6, pp. 1217-1227, June 2013.

\bibitem{Heigold2016}
G. Heigold, I. Moreno, S. Bengio, and N. Shazeer, ``End-to-end text dependent speaker verification," in \emph{Proceeding of IEEE International Conference on Acoustics, Speech, and Signal Processing (ICASSP)}, pp. 5115-5119, 2016.

\bibitem{WanLi2018}
L. Wan, Q. Wang, A. Papir and I.Moreno, ``Generalized End-to-End Loss for Speaker Verification," in \emph{Proceeding of IEEE International Conference on Acoustics, Speech and Signal Processing (ICASSP)}, pp. 4879-4883, 2018.

\bibitem{backend1}
L. Ferrer, M. Mclaren, ``A Speaker Verification Backend for Improved Calibration Performance across Varying Conditions,". in \emph{Proc. of Odyssey, the Speaker and Language Recognition Workshop}, pp. 372-379, 2020.

\bibitem{backend2}
L. Ferrer, M. McLaren, N. Brummer, ``A speaker verification backend with robust performance across conditions," 	\emph{arXiv:2102.01760}, 2021.

\bibitem{Rohdin2020}
J. Rohdin, A. Silnova, M. D{\'i}ez, O. Plchot, P. Matejka, L. Burget, O. Glembek, ``End-to-end DNN based text-independent speaker recognition for long and short utterances," Comput. Speech Lang., vol. 59, pp. 22-35, 2020.

\bibitem{neuralPLDA}
S. Ramoji, P. Krishnan, S. Ganapathy, ``Neural PLDA Modeling for End-to-End Speaker Verification," in \emph{INTERSPEECH}, 2020.

\bibitem{Lasserre2006}
A. Lasserre, C. Bishop, T. Minka, ``Principled Hybrids of Generative and Discriminative Models," in \emph{Proceeding of IEEE Computer Society Conference on Computer Vision and Pattern Recognition (CVPR)}, pp. 87-94, 2006.

\bibitem{BOSARIS2011}
N. Brummer, E. Villiers, ``The BOSARIS toolkit user guide: Theory, algorithms and code for binary classifier score processing," Documentation of BOSARIS toolkit, 2011.

\bibitem{Lehmann2005}
E. Lehmann, J Romano, Testing Statistical Hypotheses, Springer-Verlag New York, 2005.
\bibitem{Platt1999}
J. Platt, ``Probabilistic Outputs for Support Vector Machines and Comparisons to Regularized Likelihood Methods," \emph{Advances in large margin classifiers}, pp. 61-74, 1999.

\bibitem{HLin2007}
H. Lin, C. Lin, R. Weng, ``A note on Platt's probabilistic outputs for support vector machines," \emph{Machine Learning}, vol. 68, pp. 267-276, 2007.

\bibitem{LUCLS2017}
X. Lu, P. Shen, Y. Tsao, H. Kawai, ``Regularization of neural network model with distance metric learning for i-vector based spoken language identification," \emph{Computer Speech and Language}, vol.44, pp. 48-60, 2017.

\bibitem{lennorm}
D. Garcia-Romero and C. Y. Espy-Wilson, ``Analysis of i-vector length normalization in speaker recognition systems," in \emph{Proc. of INTERSPEECH}, pp. 249-252, 2011.


\bibitem{SITW2016}
M. McLaren, L. Ferrer, D. Castan, and A. Lawson, ``The speakers in the wild (SITW) speaker recognition database," in \emph{Proc. of INTERSPEECH}, pp. 818-822, 2016.

\bibitem{Vox2020}
A. Nagrani, J. Chung, W. Xie, A. Zisserman, ``Voxceleb: Large-scale speaker verification in the wild," \emph{Computer Science and Language}, vol. 60, 2020.

\bibitem{KaldiSV}
https://github.com/kaldi-asr/kaldi/tree/master/egs/voxceleb/v2

\bibitem{He2015}
K. He, X. Zhang, S. Ren, J. Sun, ``Delving Deep into Rectifiers: Surpassing Human-Level Performance on ImageNet Classification," in \emph{Proceeding of IEEE International Conference on Computer Vision (ICCV)}, pp. 1026-1034, 2015.

\bibitem{Adam}
D. P. Kingma, J. Ba, ``Adam: A Method for Stochastic Optimization," \emph{the 3rd International Conference on Learning Representations (ICLR}), 2014.

\bibitem{Xing2002}
E. Xing, A. Ng, M. Jordan, and R. Russell, ``Distance Metric Learning, with application to Clustering with side-information," \emph{in Proceeding of Advances in Neural Information Processing Systems}, MIT Press, pp. 521-528, 2002.

\bibitem{Weinberger2009}
K. Weinberger, L. Saul, ``Distance Metric Learning for Large Margin Classification," \emph{Journal of Machine Learning Research}, vol. 10, pp. 207-244, 2009.

\bibitem{BayesianFace}
B. Moghaddam, T. Jebara, A. Pentland, ``Bayesian face recognition," \emph{Pattern Recognition}, vol. 33, pp. 1771-1782, 2000.

\bibitem{ResNetSV1}
J. S. Chung, A. Nagrani, and A. Zisserman, ``Voxceleb2: Deep speaker recognition," in INTERSPEECH, 2018.

\bibitem{ResNetSV2}
D. Garcia-Romero, A. McCree, D. Snyder, and G. Sell, ``JHUHLTCOE system for the VoxSRC speaker recognition challenge," in \emph{Proc. of ICASSP}, pp. 7559-7563, 2020.

\bibitem{ResNetSV3}
H. Zeinali, S. Wang, A. Silnova, P. Matejka, and O. Plchot, ``BUT system description to VoxCeleb speaker recognition challenge 2019," arXiv:1910.12592

\bibitem{ResNetSV4}
L. Chen, K. Lee, L. He, F. Soong, ``On Early-stop Clustering for Speaker Diarization," in \emph{Proc. of Odyssey, The Speaker and Language Recognition Workshop}, pp. 110-116,  2020.

\bibitem{DAUSV1}
D. Raj, D. Snyder, D. Povey, S. Khudanpur, ``Probing the Information Encoded in X-Vectors," \emph{IEEE Automatic Speech Recognition and Understanding Workshop (ASRU)}, 2019.

\bibitem{NewSV1}
J. Deng, J. Guo, N. Xue, and S. Zafeiriou, ``ArcFace: Additive angular margin loss for deep face recognition," in \emph{IEEE/CVPR}, pp. 4685-4694, 2019.

\bibitem{NewSV2}
X. Xiang, S. Wang, H. Huang, Y. Qian, K. Yu, ``Margin Matters: Towards More Discriminative Deep Neural Network Embeddings for Speaker Recognition," in \emph{Proc. of APSIPA}, pp. 1652-1656, 2019.
\bibitem{NewSV3}
J. Chung, J. Huh, S. Mun, M. Lee, H. Heo, S. Choe, C. Ham, S. Jung, B. Lee, I. Han, ``In Defence of Metric Learning for Speaker Recognition," in \emph{Proc. of INTERSPEECH}, pp. 2977-2981, 2020.

\bibitem{ECAPA}
B. Desplanques, J. Thienpondt, and K. Demuynck, ``ECAPA-TDNN: Emphasized Channel Attention, Propagation and Aggregation in TDNN Based Speaker Verification," in \emph{Proc. INTERSPEECH}, pp. 3830-3834, 2020.

\end{thebibliography}

\end{document}